\documentclass{pasj00}

\begin{document}
\SetRunningHead{M. Honma et al.}{VERA observations of OH 43.8$-$0.1 H$_2$O masers}
\Received{2005/02/15}
\Accepted{2005/05/04}

\title{Multi-epoch VERA Observations of H$_2$O masers in OH 43.8$-$0.1}


\author{
Mareki \textsc{Honma},\altaffilmark{1,2}
Takeshi \textsc{Bushimata},\altaffilmark{1,3}
Yoon Kyung \textsc{Choi},\altaffilmark{1,4}
Takahiro \textsc{Fujii},\altaffilmark{1,5}
Tomoya \textsc{Hirota},\altaffilmark{1}\\
Koji \textsc{Horiai},\altaffilmark{1,6}
Hiroshi \textsc{Imai},\altaffilmark{5}
Noritomo \textsc{Inomata},\altaffilmark{5}
Jose \textsc{Ishitsuka},\altaffilmark{1}
Kenzaburo \textsc{Iwadate},\altaffilmark{1,6}\\
Takaaki \textsc{Jike},\altaffilmark{1}
Osamu \textsc{Kameya},\altaffilmark{1,2,6}
Ryuichi \textsc{Kamohara},\altaffilmark{1,5}
Yukitoshi \textsc{Kan-ya},\altaffilmark{1}\\
Noriyuki \textsc{Kawaguchi},\altaffilmark{1,2,3}
Masachika \textsc{Kijima},\altaffilmark{5}
Hideyuki \textsc{Kobayashi},\altaffilmark{1,3,4}
Seisuke \textsc{Kuji},\altaffilmark{1,6}\\
Tomoharu \textsc{Kurayama},\altaffilmark{1,4}
Seiji \textsc{Manabe},\altaffilmark{1,2,6}
Takeshi \textsc{Miyaji},\altaffilmark{1,3}
Akiharu \textsc{Nakagawa},\altaffilmark{5}\\
Kouichirou \textsc{Nakashima},\altaffilmark{5}
Chung Sik \textsc{Oh},\altaffilmark{1,4}
Toshihiro \textsc{Omodaka},\altaffilmark{5}
Tomoaki \textsc{Oyama},\altaffilmark{1,4}
Maria \textsc{Rioja},\altaffilmark{7}\\
Satoshi \textsc{Sakai},\altaffilmark{1,6}
Katsuhisa \textsc{Sato},\altaffilmark{1,6}
Tetsuo \textsc{Sasao},\altaffilmark{8}
Katsunori M. \textsc{Shibata},\altaffilmark{1,3}
Rie \textsc{Shimizu},\altaffilmark{5}\\
Kasumi \textsc{Sora},\altaffilmark{5}
Hiroshi \textsc{Suda},\altaffilmark{1,4}
Yoshiaki \textsc{Tamura},\altaffilmark{1,2,6} and 
Kazuyoshi \textsc{Yamashita}\altaffilmark{5}
}

\altaffiltext{1}{VERA Observatory, NAOJ, Mitaka, Tokyo 181-8588}
\altaffiltext{2}{Graduate University for Advanced Studies, Mitaka, Tokyo 181-8588}
\altaffiltext{3}{Space VLBI Project, NAOJ, Mitaka, Tokyo 181-8588}
\altaffiltext{4}{Department of Astronomy, University of Tokyo, Bunkyo, Tokyo 113-8654}
\altaffiltext{5}{Faculty of Science, Kagoshima University, Korimoto, Kagoshima, Kagoshima 890-0065}
\altaffiltext{6}{Mizusawa Astrogeodymaics Observatory, NAOJ, Mizusawa, Iwate 023-0861}
\altaffiltext{7}{Observatorio Astron\'{o}mico Nacional, Madrid, Spain}
\altaffiltext{8}{Ajou University, Suwon 442-749, Republic of Korea}

\email{honmamr@cc.nao.ac.jp}


\maketitle

\begin{abstract}

We report on the multi-epoch observations of H$_2$O maser emission in star forming region OH 43.8$-$0.1 carried out with VLBI Exploration of Radio Astrometry (VERA).
The large-scale maser distributions obtained by single-beam VLBI mapping reveal new maser spots scattered in area of 0.7 x 1.0 arcsec, in addition to a {`}shell-like{'} structure with a scale of 0.3 x 0.5 arcsec which was previously mapped by Downes et al.(1979).
Proper motions are also obtained for 43 spots based on 5-epoch monitoring with a time span of 281 days.
The distributions of proper motions show a systematic outflow in the north-south direction with an expansion velocity of $\sim$ 8 km s$^{-1}$, and overall distributions of maser spots as well as proper motions are better represented by a bipolar flow plus a central maser cluster with a complex structure, rather than a shell with a uniform expansion such as those found in Cep A R5 and W75N VLA2.
The distance to OH 43.8$-$0.1 is also estimated based on the statistical parallax, yielding $D = 2.8$ $\pm$ 0.5 kpc.
This distance is consistent with a near kinematic distance and rules out a far kinematics distance ($\sim$ 9 kpc), and the LSR velocity of OH 43.8$-$0.1 combined with the distance provides a constraint on the flatness of the galactic rotation curve, that there is no systematic difference in rotation speeds at the Sun and at the position of OH 43.8$-$0.1, which is located at the galacto-centric radius of $\sim$ 6.3 kpc.

\end{abstract}

\section{Introduction}

Star forming regions are often associated with water maser emission, and multi-epoch VLBI observations of water masers provide us a unique tool to study the structure and kinematics of star forming regions through proper motions of maser spots.
Until the middle of 1990s, VLBI studies of H$_2$O maser sources were made to reveal the kinematics in the well-known star forming regions such as Orion KL, W51M\&N, Sgr B2, W49N, W3(OH) (Genzel et al. 1981a; 1981b; Schneps et al. 1981; Reid et al. 1988; Gwinn et al. 1992; Alcolea et al. 1993).
These studies successfully detected proper motions of maser spots, and demonstrated that maser-emitting gases trace the shock regions where the outflows from forming stars hit ambient gases.
Proper motions of maser spots were also used to determine the source distance via statistical parallax and/or detailed kinematic modeling, providing a powerful tool to determine distances in galactic scale, even up to $\sim$ 10 kpc.
Recently, numerous efforts were made to perform VLBI studies of more number of star forming regions, detecting proper motions of H$_2$O masers in several other sources such as IRAS 05413-0104, IRAS 21391+5802, S106 FIR, W3 IRS5, W51S, Cep A, W75N, NGC 2071, AFGL 5142, IRAS 20050+2720 (e.g., Claussen et al. 1998; Patel et al. 2000; Furuya et al. 2000; Imai et al. 2000; 2002; Torrelles et al. 2001; 2003; Seth et al. 2002; Goddi et al. 2004; Furuya et al. 2005).
While most of H$_2$O maser sources show bipolarity in their distributions as well as motions, there are a few exceptions exhibiting coherently expanding shell (Cep A R5 \& W75N VLA2: Torrelles et al. 2001; 2003), casting a question on the formation mechanism of these morphologically different outflows in star forming regions.

OH 43.8$-$0.1 (also referred to as G 43.8-0.1 or G 43.795-0.127) is a star forming region located in the galactic plane with OH, CH$_3$OH and H$_2$O maser emissions.
While optical and near IR observations tell little about this region due to the extremely large galactic extinction, this star forming region is known to be associated with an infrared source IRAS 19095+0930 and an Ultra-Compact HII region (hereafter UC HII; Kurtz et al. 1994).
VLA studies of OH maser distributions revealed that OH maser spots are surrounding the UC HII (Argon et al. 2000).
Molecular lines such as CO and CS were also detected toward the direction of OH 43.8$-$0.1 (Plume et al. 1992; Shepherd \& Churchwell 1996), providing a systemic velocity of $V_{\rm LSR}\sim 44$ km s$^{-1}$.
Assuming the standard galactic rotation (a flat rotation curve with $R_0=8.5$ kpc and $\Theta_0=220$ km s$^{-1}$), the systemic velocity gives a near kinematic distance of 3 kpc and a far distance of 9 kpc.
H$_2$O maser emission in OH 43.8$-$0.1 was discovered by Genzel \& Downes (1977) and Dickinson et al.(1978), and shortly later VLBI observations of H$_2$O maser were carried out by Downes et al.(1979).
The distribution of maser spots obtained by Downes et al. (1979) suggested a shell-like structure with a diameter of 0.5 arcsec.
Therefore, OH 43.8$-$0.1 is another candidate for H$_2$O maser sources that show shell-like morphology such as those recently found in Cep A R5 and W75N VLA2 (Torrelles et al. 2001; 2003).
In order to confirm (or rule out) this, new multi-epoch VLBI observations are necessary to measure the proper motions of maser spots.

Here we present results of multi-epoch observations OH 43.8$-$0.1 H$_2$O masers with VLBI Exploration of Radio Astrometry (VERA), which is a new Japanese VLBI array dedicated to phase-referencing VLBI astrometry (Honma et al. 2000; Kobayashi et al. 2003).
Originally the observations of OH 43.8$-$0.1 were carried out as a series of test observations to evaluate the system performance of VERA, by observing OH 43.8$-$0.1 and another bright maser source, W49N, simultaneously in the dual-beam mode.
The series of test observations already demonstrated a high capability of phase-referencing with VERA's dual-beam system (Honma et al. 2003), and also produced maser maps of W49N, successfully identifying the outburst spots which flared up in 2003 October (Honma et al. 2004).
In the present paper, we focus on the results on OH 43.8$-$0.1, analyzed as single-beam VLBI observations.
We present the maser spots map in larger scale than that of Downes et al.(1979), and measure proper motions for the first time for OH 43.8$-$0.1 based on 5-epoch observations spanning 281 days.
We also determine the distance to OH 43.8$-$0.1 based on statistical parallax, and discuss the structure and kinematics of OH 43.8$-$0.1 maser as well as constraint on galactic rotation curve obtained from the distance to OH 43.8$-$0.1.

\section{Observations and Reductions}

VERA observations of OH 43.8$-$0.1 were performed for 9 epochs between October in 2003 and September in 2004 with typical intervals of 1 $\sim$ 2 months as a series of test observations of VERA.
The observations were done in dual-beam mode observing two bright H$_2$O maser sources W49N and OH 43.8$-$0.1 (0$^\circ$.65 separation) simultaneously to evaluate the capability of phase-referencing as well as astrometric precision.
The results of phase-referencing performance tests as well as dual-beam astrometric capability will be discussed elsewhere, and here we concentrate on OH 43.8$-$0.1.
Left-hand circular polarization signals were digitized and recorded using the VSOP-terminal system at a data rate of 128 Mbps with 2-bit quantization.
Among the total bandwidth of 32 MHz, one 16 MHz channel was assigned to OH 43.8$-$0.1, covering the radial velocity span of 215.7 km s$^{-1}$ (adopting the rest frequency of 22.235080 GHz for H$_2$O 6$_{16}$-5$_{23}$ transition).
Correlation processing was performed with the Mitaka FX correlator with a spectral resolution of 512 points per channel, yielding frequency and velocity resolutions of 31.25 kHz and 0.42 km s$^{-1}$, respectively.

In some epochs, one or a few stations were partly or fully missed due to system trouble and/or bad weather.
Since VERA array has only 4 stations (Mizusawa, Iriki, Ogasawara, and Ishigaki-jima), lack of a station causes severe degrading of synthesized images, and would introduce a large error in doing multi-epoch measurements of maser proper motions.
Hence, in the present paper, we use the data of 5 epochs that were obtained with the full 4-station array under relatively good conditions.
The epochs presented here are: day of year (DOY) 356 in 2003, DOY 054, 120, 205, and 272 in 2004 (Dec. 22 in 2003, Feb. 23, Apr. 29, Jun. 23, and Sep. 28 in 2004).
The system noise temperatures at the zenith were typically 150 K to 250 K for the first three epochs, and 200 K to 300 K for the last two epochs.

In the data analysis of each epoch, calibrations of clock parameters, bandpass, amplitude and phase were carried out in a standard manner.
Clock parameters (clock offset and clock rate offset) were calibrated using the residual delay and delay rate for a bright calibrator source, TXS 1923+210 (= ICRF J192559.6+210626), which was observed every 2 hours.
The amplitude calibrations were done using the system noise temperature, which were evaluated by the {`}R-Sky{'} method, observing the reference black body at the beginning of each scan (typically every hour).
Bandpass calibrations were done using auto-correlation spectra of continuum sources.
For phase calibrations, visibilities of all velocity channels were phase-referenced to the reference maser spot at $V_{\rm LSR}$ of 38.2 km s$^{-1}$, which is one of the brightest spots, and shows no sign of structure according to the closure phase.
After those processes, calibrated visibilities were Fourier transformed to synthesized images, and the positions of the brightness peaks were determined with respect to the reference maser spot (note that the reference spot was the same for all epochs).
The synthesized beam has a FWHM beam size of 1.2$\times$2.3 mas with a PA of 45$^\circ$.
In order to detect maser spots that are far from the reference spot, spot positions were first estimated using fringe rates, and the detailed mapping is performed around the estimated positions.

For analyses of multi-epoch VLBI data, we identified maser spots based on spot positions as well as radial velocities.
Spots in different epochs were identified as {`}same{'} if their positions were within 1 mas and their radial velocities were within 0.21 km s$^{-1}$, which is a half of velocity resolution.
For maser spots that were detected for more than 3 epochs, their proper motions were determined based on the linear least-squares-fit.

\section{Results}
\subsection{Spectral Evolution}

Figure 1 shows the spectral evolutions of OH 43.8$-$0.1 for the 5 epochs.
The LSR velocities of maser lines range from 26 to 53 km s$^{-1}$, showing no major difference from the observations in 1977 (Downes et al. 1979).
As seen in figure 1, auto-correlation spectra of OH 43.8$-$0.1 showed rapid variability of maser intensity during our observations.
The strong variability of OH 43.8$-$0.1 was already found by long-term monitoring by Lekht (2000), and similar variabilities are found in many of H$_2$O maser sources, including rapid flare-up of a certain feature.
The most prominent variation occurred at $V_{\rm LSR}$ of 38.2 km s$^{-1}$, which had 1.0$\times 10^3$ Jy on DOY 356 in 2003 and decreased to 75 Jy on DOY of 272 in 2004.
Lekht (2000) also found a rapid flare at $V_{\rm LSR}=$38.2 km s$^{-1}$ in 1997 exceeding 3.7$\times 10^3$ Jy with a duration of 6 months.

Figure 1 also shows the cross-correlated spectra of OH 43.8$-$0.1 for Mizusawa---Ishigaki-jima baseline, which is the longest among the 6 baselines of VERA.
The cross-correlation spectra in figure 1 were taken around the maximum UV distance, being $\sim$ 2200 km or 1.6$\times 10^8 \lambda$.
In figure 1, for some features correlated flux is much smaller than total power flux, indicating that maser features are partially or fully resolved out with this baseline.
The LSR velocities of cross-correlated maser lines mostly range from 34 to 51 km s$^{-1}$, while some features around 25 to 34 km s$^{-1}$ are detected in cross-correlated spectrum on DOY 120 in 2004.
The component of $V_{\rm LSR}=$38.2 km s$^{-1}$ is strongest in cross-correlated spectra in the first three epochs, with its peak intensity of 870 Jy on DOY 356 in 2003.
The high ratio of correlated flux to total flux (8.7$\times 10^2$ Jy to 1.0$\times 10^3$ Jy) suggests that the flaring of the 38.2 km s$^{-1}$ spot occurred in a compact area, being smaller than 6 AU (corresponding to 2 mas at the distance of 2.8 kpc; see subsection 3.4 for distance determination). 
In the following analyses, the maser spot at 38.2 km s$^{-1}$ was taken to be the reference spot for visibility phase calibration.

\begin{figure}
\begin{center}
	\includegraphics[clip,width=11.7cm,angle=-90]{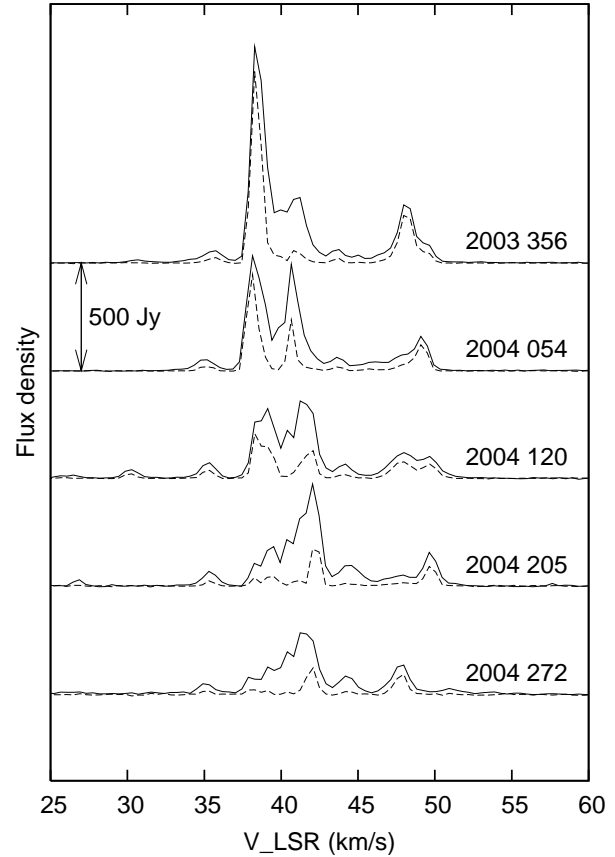}
\end{center}
\caption{Five-epoch spectral evolution of OH 43.8$-$0.1 H$_2$O masers. Thin lines show total flux (auto-correlation), and dashed lines show cross-correlated flux for Mizusawa---Ishigaki-jima baseline. The vertical arrow shows the flux scale of 500 Jy, and each spectrum is displaced by the same amount.}
\end{figure}

\subsection{Maser Spot Distributions}

Figure 2 (a) shows the composite map of OH 43.8$-$0.1 maser spot distributions with a size of $1^{"}\!\!.5$ $\times$ $1^{"}\!\!.5$.
Spots that are detected at least in one epoch are plotted.
The number of detected maser spots in each epoch are 53, 54, 60, 47 and 51 from DOY 356 in 2003 to DOY 272 in 2004.
For masers spots detected for more than 3 epochs, proper motion vectors are superposed (for proper motion measurements, see next subsection).

The map origin of figure 2 (a) is the position of the reference maser spot at $V_{\rm LSR}=$38.2 km s$^{-1}$.
The absolute position of the map center is estimated to be ($\alpha_{J2000}$, $\delta_{J2000}$)=($19^{h}11^{m}53^{s}\!\!.986$, $+09^\circ 35^{'}50^{"}\!\!.22$) with an uncertainty of $0^{''}\!\!.2$ using the fringe rates of the reference maser spot.
Color index in figure 2 shows the LSR velocity range from 34.4 km s$^{-1}$ to 50.9 km s$^{-1}$, where most of maser spots are located.
Only exceptions are those with LSR velocities of 25 $\sim$ 34 km s$^{-1}$, which were only detected on DOY 120 of 2004 and shown as open circles in figure 2 (b) and 2 (c).

Figure 2 (b) is an expanded map of the shell-like structure discovered by Downes et al. (1979).
While figure 2 (b) is consistent with the map obtained by Downes et al., figure 2 (a) shows that the shell-like structure found previously do not represent the entire region of maser spot distributions.
In fact, in figure 2 (a), the maser distributions show an extension with a scale of $\sim$ 1 arcsec from the northern top of {`}shell-like{'} structure toward the south.
This extension of maser spots were not reported in previous studies and discovered for the first time in the present study.
The northern maser spots at (-145, 136) have $V_{\rm LSR}$ of 47 to 51 km s$^{-1}$, and the southern-most maser spot at (129, -810) has $V_{\rm LSR}$ of 41 km s$^{-1}$.
These properties of spot positions as well as radial velocities indicate that the maser spot distribution in OH 43.8$-$0.1 is not represented by a simple shell, but may be interpreted as a bipolar structure in the north-south direction.

\begin{figure}[th]
\begin{center}
	\includegraphics[clip,width=13.5cm,angle=-90]{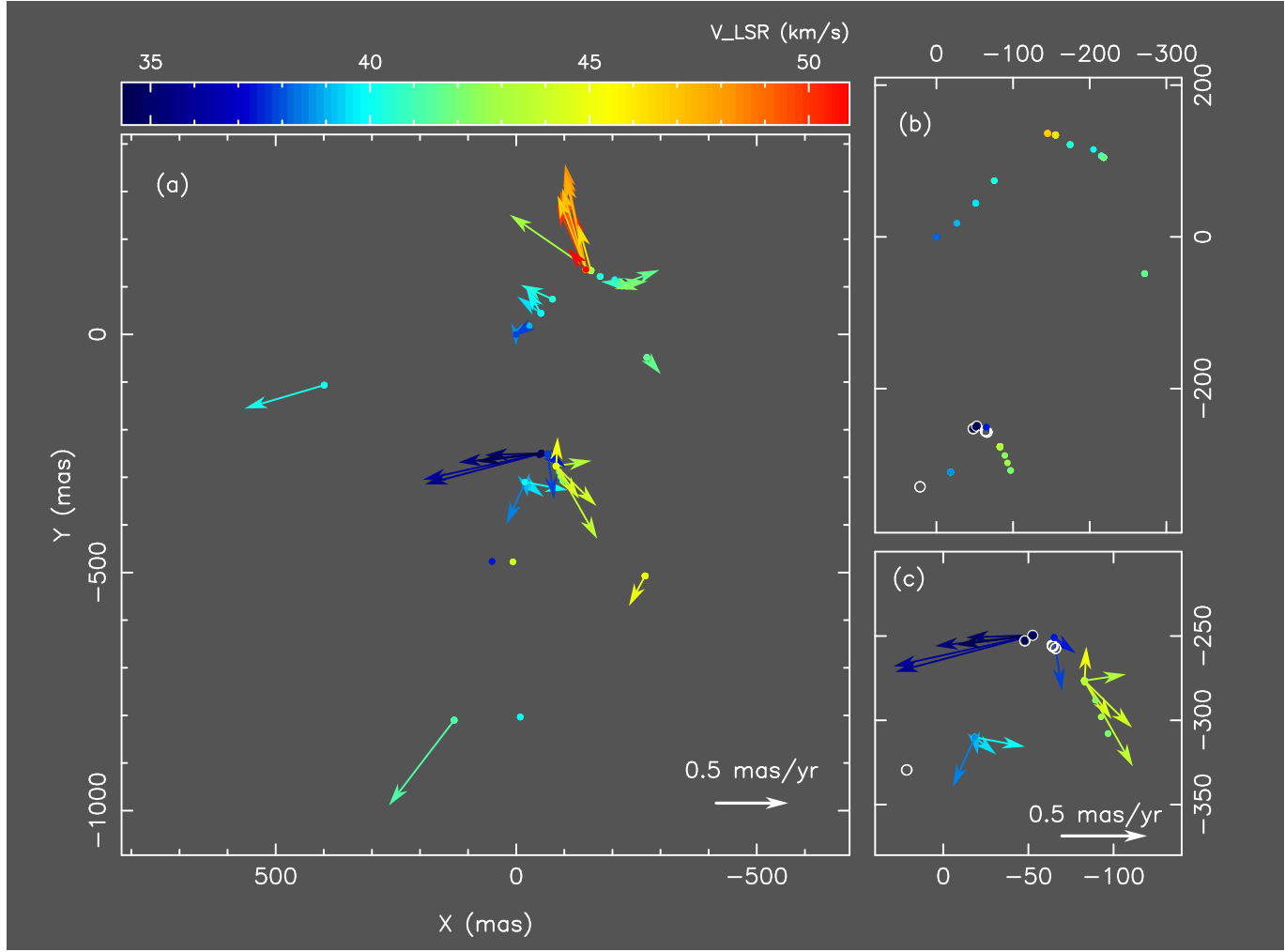}
\end{center}
\caption{Composite map of maser spot distributions and proper motion vectors of OH 43.8$-$0.1. Spot color indicates the LSR velocity (see color index at the top). 
The systemic velocity of OH 43.8$-$0.1 is $V_{\rm LSR}\approx 42$ km s$^{-1}$.
Left (a): the large scale map with a size of 1.5 arcsec $\times$ 1.5 arcsec (1 arcsec corresponds to 2800 AU at the distance of 2.8 kpc). 
The arrow at bottom-right corner indicates the proper motion of 0.5 mas yr$^{-1}$, corresponding to 6.6 km s$^{-1}$ at the distance of 2.8 kpc.
Top-right (b): expansions of the shell-like structure found by Downes et al.(1979). Bottom-right (c): close-up to the central cluster with proper motion vectors. Circles in (b) and (c) are maser spots with radial velocities out of the color index range ($V_{\rm LSR}$=25 to 34 km s$^{-1}$) detected only in DOY 120 of 2004.
}
\end{figure}

\noindent
In addition to the north-south extension, there are a few other maser spots which are rather isolated, such as the one at ($X$,$Y$)=(399, -107) and (-268, -507), which were also not reported in the previous studies.

\vspace*{17cm}
\noindent

\subsection{Proper Motion Measurements}

As described in the previous section, maser spots in different epochs were identified as {`}same{'} if their positions and velocities are within 1 mas and 0.21 km s$^{-1}$, respectively.
Based on these identification criteria, 43 maser spots were identified in more than 3 epochs, and proper motions were determined based on the linear least-squares-fit.
Table 1 summarizes the results of proper motion measurements obtained by a linear least-square-fit, including LSR velocities, spot positions, proper motions, detected epochs, and RMS residuals.
Note that the RMS residuals $\delta_X$ and $\delta_Y$ are defined as $\delta_X=\sqrt{\Sigma(X_{\rm obs}-X_{\rm fit})^2 /n}$, where $X_{\rm obs}$ and $X_{\rm fit}$ are observed and best-fit positions of spots and $n$ is the number of data ($3\leqq n\leqq 5$).
Values of $\delta_X$ and $\delta_Y$ provide us uncertainty estimates in the proper motion measurements.
The maximum values of $\delta_X$ and $\delta_Y$ are 0.060 and 0.082 mas, and the mean values of $\delta_X$ and $\delta_Y$ for 43 spots are 0.022 and 0.029 mas.
These values indicate that the proper motion measurements were successfully done with an accuracy better than 0.1 mas.
Also, in table 1 there is no systematic trend in $\delta_X$ and $\delta_Y$ with LSR velocity or positions of maser spots.
Since the maser spots are bright enough (typically larger than 10 Jy in correlated flux), the residuals from the best-fit linear motion are likely to be dominated by internal structure of maser spots rather than thermal noise.
In fact, adopting the distance of 2.8 kpc (see next subsection for distance determination), the mean value of $\delta_X$ and $\delta_Y$ correspond to 0.06 AU and 0.08 AU.
Observations of other H$_2$O maser sources suggested that maser clouds have internal structure with a typical scale of 1 AU and a typical velocity width of 0.5 km s$^{-1}$ (e.g., Gwinn 1994).
Assuming the same values for masers in OH 43.8$-$0.1, one can roughly estimate possible changes in maser peak positions due to internal turbulent motions during our time span of observations (281 days) as 0.5 km s$^{-1} \times 281 \times 86400$ s = 0.08 AU, which is quite close to the physical scales given by the mean values of $\delta_X$ and $\delta_Y$.

The proper motion vectors of maser spots listed in table 1 are shown in figure 2 (a).
Proper motion vectors are shown for the spots with proper motions greater than 0.1 mas yr$^{-1}$ (40 among the 43 spots in table 1).
As described in the previous section, the maser spot at 38.2 km s$^{-1}$ was taken to be the reference maser spot, and thus all the proper motion vectors are relative to the proper motion of the reference spot itself.
The proper motion of the reference maser spot could not be measured in the present study, and thus it might introduce a bias to proper motions of other maser spots if the reference maser spot itself has a large proper motion.
However, the mean values of $\mu_X$ and $\mu_Y$ for 43 spots in table 1 are 0.117 mas yr$^{-1}$ and 0.086 mas yr$^{-1}$, corresponding to 1.6 km s$^{-1}$ and 1.1 km s$^{-1}$ at the source distance of 2.8 kpc.
These values of mean proper motions are fairly smaller than the radial velocity width ($\sim$25 km s$^{-1}$, see figure 1) and typical proper motion in table 1 ($\sim$0.7 mas yr$^{-1}$), indicating that the proper motion of the reference maser spot does not introduce large systematic motions of the reference frame in figure 2.

As seen in figure 2 (a), the global distribution of proper motions show bipolar structure with red-shifted northern components around ($X$, $Y$)=(-145, 136) moving toward the north and moderately blue-shifted southern component at (129, -810) moving toward the south.
Note that while there are more than 10 spots in the red-shifted northern components, there is only one blue-shifted southern component for which proper motion was measured.
This is likely to be due to the geometric effect of bipolar flow which causes the difference in the column density of masing gas toward these components: red-shifted components locate on the far side of the protostar along the line-of-sight and thus the column density of masing gas is higher than the blue-shifted components, which locates on the near side of the protostar.
Between these two components there is a cluster of maser spots with $X$ between -80 and -20 and $Y$ between -310 and -250, which hereafter we refer to as the {`}central cluster{'}, shown in figure 2 (c).
The directions and amplitudes of proper motion vectors in the central cluster are quite different from those of northern and southern components, showing rather complex structures.
For instance, maser features at ($X$, $Y$)=(-83, -276) (spot number 13, 15 to 18 in table 1) and at (-18, -310) (spot number 29, 31 to 33) exhibit large velocity changes within the maser features in 0.3 mas scale ($\sim$ 0.8 AU).
Highly blue-shifted maser spots at radial velocity of 25 to 34 km s$^{-1}$ (which were detected only for DOY 120, 2004) also locate in the central cluster area (open circles in figure 2 c).
These complex structure of the central cluster may indicate that the protostar is associated with the central cluster, and the protostar's activity causes the complexity.

In contrast to an expanding shell such as those found in Cep A R5 (Torrelles et al. 2001) or W75N VLA2 (Torrelles et al. 2003), the proper motions in the shell-like structure found by Downes et al. cannot be represented by a coherent expansion from the shell center.
The proper motions of the {`}shell{'} in the north-south direction are 0.5 to 0.7 mas yr$^{-1}$, and are much larger than those of the east-west direction, which are around 0.1 to 0.2 mas yr$^{-1}$.
The global properties of proper motion distributions indicate that the structure of OH 43.8$-$0.1 H$_2$O maser can be represented by bipolar flow plus central cluster, and that the shell-like structure in figure 2 (b) is just a part of bipolar flow and the central cluster (for further discussion on the structure of OH 43.8$-$0.1, see next section).

\subsection{Distance Determinations by Statistical Parallax}

Distances to maser sources can be determined from the dispersions of radial velocities and proper motions based on statistical parallax.
Statistical parallax assumes only that the velocity dispersions are isotropic, and the basic equation for distance determinations with statistical parallax can be written as,
\begin{equation}
D = \frac{\sigma_V}{\sigma_{\mu}} \equiv \frac{\sigma_V}{\sqrt{(\sigma_{\mu X}^2 + \sigma_{\mu Y}^2)/2}}.
\label{eq:D}
\end{equation}
Here $D$ is the source distance, $\sigma_V$ is the radial velocity dispersion, and $\sigma_{\mu}$ is the one-dimensional proper motion dispersion, and $\sigma_{\mu X}$ and $\sigma_{\mu Y}$ are the proper motion dispersions in $X$ and $Y$ directions, respectively.
In practice, however, equation (\ref{eq:D}) is too much simplified, and one has to take into account the effect of observational uncertainties using following equation instead of equation (\ref{eq:D}),
\begin{equation}
D = \frac{\sigma_V'}{\sigma_{\mu}'} \equiv \frac{\sqrt{\sigma_V^2 - \epsilon_V^2}}{\sqrt{\left[(\sigma_{\mu X}^2 - \epsilon_X^2) + (\sigma_{\mu Y}^2 - \epsilon_Y^2)\right]/2}}.
\label{eq:D_eps}
\end{equation}
The $\epsilon$ terms correct for the effect of observational uncertainty that broadens the widths of velocity and proper motion.
In the following analyses we make use of equation (\ref{eq:D_eps}) as the basic equation for distance determination based on statistic parallax (but note that equations (\ref{eq:D}) and (\ref{eq:D_eps}) give almost identical results because the $\epsilon$ terms are sufficiently small in the present study).

In several cases H$_2$O maser sources show bipolarity and thus anisotropy, and so careful analyses are necessary to obtain reliable distances.
In the present paper, we perform statistical parallax analyses based on four different definitions of $\sigma_{\mu X}$ and $\sigma_{\mu Y}$: 1) using all maser spots with dispersion $\sigma$ defined as root-mean-square value from the average (i.e., $\sigma_{\mu X}=\sqrt{\Sigma (\bar{\mu}_X-\mu_{X,i})^2/N}$), 2) using all maser spots with dispersion $\sigma$ defined as the peak-to-peak value (i.e., $\sigma_{\mu X}=\max(\mu_{Xi})-\min(\mu_{Xi})$, 3) the same definition of $\sigma$ to 1) but only with spots in the central cluster, and 4) the same definition of $\sigma$ to 2) but only with spots in the central cluster.
Note that in all cases, the dispersions of LSR velocities are also defined in the same manner with those of proper motions (i.e., $\sigma_V=\sqrt{\Sigma (\bar{V}-V_i)^2/N}$ for cases 1) and 3), and $\sigma_V=\max(V_i)-\min(V_i)$ for cases 2) and 4).
Cases 3) and 4) are to avoid the anisotropy caused by the bipolarity of maser flow: as already seen in the previous sections, while the large scale proper motions are represented by a bipolar flow, the proper motions in the central cluster show rather complex structure.
Also, cases 2) and 4) are for testing how different definitions of {`}dispersion{'} would affect the results, as the distributions of LSR velocities and proper motions are not necessarily modeled by a Gaussian distribution.
In all cases, $\epsilon_V$ is taken to be 0.21 km s$^{-1}$, which is the half of velocity resolution and the threshold for identifying maser spots in different epochs, and $\epsilon_X$ and $\epsilon_Y$ are taken to be 0.029 and 0.038 mas yr$^{-1}$, which are determined from averaged values of $\delta_X$ and $\delta_Y$ and the time span $\tau$=281 day=0.77 yr, as $\epsilon_X=\bar{\delta}_X/\tau$ and $\epsilon_Y=\bar{\delta}_Y/\tau$.

Table 2 summarizes the results of distance determination with case 1) to 4).
Calculated distances vary from 3.3 kpc for case 1) to 2.3 kpc for case 4).
There are also some systematic trends in 4 cases: analyses with all spots tend to give larger distances, and analyses with RMS definition of dispersions tend to give larger distances.
The former tendency may reflect a geometric effect of bipolarity.
For instance, in cases 1) and 3) dispersions of proper motions are similar, but dispersions of LSR velocities are increased by factor of 1.26 when all spots are included (case 1), yielding the farther distance.
This result may imply that the location of the bipolar structure is head-on, i.e., the angle between bipolar axis and line-of-sight is less than 45 degrees.
The origin of latter tendency in table 2 (larger distance for RMS definition of dispersions) is unclear, but the difference of distances for two definitions of $\sigma$ is a factor of 1.14, and this could be an error introduced by small number statistics ($N$=16 or 43).

It is not easy to judge which of the four cases best represents the true distance to OH 43.8$-$0.1, as all four cases have advantages and disadvantages.
While analyses with spots only in the central cluster avoid the effect of bipolarity, the number of spots become smaller and degrade statistic reliability.
The peak-to-peak definition of dispersions is severely affected by an outlier with an extreme LSR velocity or proper motions (if exists), while the RMS definitions of dispersions may not work well for non-Gaussian distributions (e.g., $V_{\rm LSR}$ distribution in figure 1).
In order to obtain most-likely value of the distance to OH 43.8$-$0.1, here we conservatively take an average of 4 cases, which yields $D = 2.8 \pm 0.5$ kpc.
This distance is consistent with the near kinematics distance ($\sim 3$ kpc), and rules out the far kinematic distance ($\sim 9$ kpc).

\begin{table}
Table~2. \hspace{4pt} Results of distance determinations
\vspace{6pt}\\
\begin{tabular}{ccccccc}
\hline
 case & spot$\;^{\rm a}$ & $N\;^{\rm b}$ & def $\sigma$ $^{\rm c}$ & $\sigma_V'\;^{\rm d}$ & $\sigma_{\mu}'\;^{\rm e}$ & $D\;^{\rm f}$ \\
\hline
1 & all & 43 & RMS & 4.69 & 0.303 & 3.28 \\
2 & all & 43 & PP & 16.5 & 1.23 & 2.82 \\
3 & CC & 16 & RMS & 3.72 & 0.294 & 2.67 \\
4 & CC & 16 & PP & 10.1 & 0.930 & 2.30 \\
\hline
 & & & & & mean & 2.8$\;\pm\;$0.5 \\
\hline
\end{tabular}
\vspace{1mm}
\\
a: spot selection: all for all spots, and CC for spots in the central cluster.\\
b: number of spots used in analyses \\
c: definitions for $\sigma_V$, $\sigma_{\mu X}$, and $\sigma_{\mu Y}$: RMS for root-mean-square, and PP for peak-to-peak definitions (see text).\\
d: $\sigma_V'$ in km s$^{-1}$ (equation [\ref{eq:D_eps}]).\\
e: $\sigma_{\mu}'$ in mas yr$^{-1}$ (equation [\ref{eq:D_eps}]).\\
f: determined distance in kpc.\\
\end{table}

\section{Discussion}

\subsection{Structure of OH 43.8$-$0.1 H$_2$O masers}

The key to understand the structure and kinematics of H$_2$O maser in the context of star formation in OH 43.8$-$0.1 is to know the location of the forming star relative to maser spots in figure 2.
Unfortunately H$_2$O maser spots are most likely to be located in shock regions where outflow from a forming star hit ambient gases, and thus the position of the forming star cannot be directly obtained from H$_2$O maser observations.
An approach to obtain the position of a forming star is to use the continuum radio emissions, as a forming OB star radiate strong UV photons and produce a compact HII region.
In fact, an Ultra-Compact HII region (UC HII) was found by VLA observations at frequencies of 8 GHz and 15 GHz (Kurtz et al. 1994).
They found an unresolved UCHII at the position of ($\alpha_{J2000}$, $\delta_{J2000}$)=($19^{h}11^{m}53^{s}\!\!.987$, $+09^\circ 35^{'}50^{"}\!\!.308$) with an uncertainty of $0^{"}\!\!.1$.
This agrees well with our absolute position determination for the reference spot, ($19^{h}11^{m}53^{s}\!\!.986$, $+09^\circ 35^{'}50^{"}\!\!.22$) $\pm 0^{"}\!\!.2$.
Thus, the UC HII regions indeed locate in the maser emitting regions, but precise location in figure 2 (a) is still unclear due to position error.

As for OH maser emissions, Argon et al.(2000) located OH maser spots with respect to UC HII region using VLA A-Array.
Their results demonstrated the clear association of the OH maser with the UC HII region.
It is notable that the LSR velocity of OH maser emissions ranges from 40 to 45 km s$^{-1}$ (Argon et al. 2000), which is fairly close to LSR velocity range of the central cluster shown in figure 2 (c) (34 to 45 km s$^{-1}$).
The results of OH maser observations by Argon et al. (2000) may imply a connection between OH maser and the central cluster of H$_2$O maser, and thus a possible link between  the central cluster and the UC HII region.

In addition to OH maser, methanol (CH$_3$OH) masers at 6.7 GHz were found in OH 43.8$-$0.1 (Menten 1991, Szymczak et al. 2000).
The 6.7 GHz methanol maser is categorized as class II methanol maser, which shows a clear connection with star-formation, presumably associated with accreting matter or accretion disk around a protostar.
Menten (1991) reported that the methanol maser line in OH 43.8$-$0.1 has $V_{\rm LSR}$ range of 39 to 44 km s$^{-1}$.
Szymczak et al. (2000) also confirmed this result, while their results show that the peak intensity occurs at 39 to 40 km s$^{-1}$.
These $V_{\rm LSR}$ ranges are closer to those of the central cluster (34 to 45 km s$^{-1}$), rather than northern maser spots at ($X$, $Y$)=(-145, 136) (46 to 51 km s$^{-1}$).
Thus, a simple interpretation of $V_{\rm LSR}$ of methanol maser lines is that methanol maser emissions are associated with the central cluster.
If these methanol emissions are also associated with a forming star and thus UC HII, this indicates that the unresolved UC HII is located in or near the central cluster.

The other way to estimate the position of forming star with respect to maser features is to use kinematic information.
Since the outflow is powered by the forming star, its position can be estimated as an origin of outflow motion.
Detailed modeling have been done for several sources to evaluate the position of outflow origin as well as distances (e.g., Reid et al. 1988; Gwinn et al. 1992; Imai et al. 2000).
Here we take one of the simplest (and often used) kinematic models which assumes a radial outflow from a single origin with a constant outflow velocity.
Basic technique is essentially the same as the one used in Imai et al.(2000), where the outflow origin was determined to minimize the differences between observed velocity vectors and model velocity vectors.
Based on this kinematic modeling, the outflow origin was estimated to be at ($X$, $Y$)=(-260 $\pm$ 130, -250 $\pm$ 130) mas.
Although the error bars are relatively large, the estimated position of the outflow origin coincides with the position of the central cluster within the error bars.
Considering the discussions described above, one can infer that the central maser cluster is associated with a forming star and the bipolar flow in the north-south direction is powered by the forming star in or near the central cluster.
This kind of structure, a bipolar flow plus a central cluster, is also seen in other H$_2$O maser sources, such as IRAS 05413-0104 (Claussen et al. 1998) and IRAS 21391+5802 (Patel et al. 2000), which are low mass star forming regions.
To obtain more firm conclusions on the case for OH 43.8$-$0.1, precise collocation of H$_2$O maser spots with UC HII regions is definitely required.

The distance to OH 43.8$-$0.1 provide us a physical scale of H$_2$O emitting regions.
The north-south bipolar extension (0.98 arcsec) corresponds to a physical scale of 2740 AU at the distance of 2.8 kpc.
This scale is remarkably smaller than the maser structures of other massive star forming regions such as W49N ($\sim 20000$ AU, Gwinn et al. 1992), Sgr B2(N) ($\sim 28000$ AU, Reid et al. 1988), and Orion-KL ($\sim 19000$ AU, Genzel et al. 1981a).
The size difference mainly originate from the small outflow velocity of OH 43.8$-$0.1 rather than age differences in these star forming regions.
In fact, dynamical age estimate of OH 43.8$-$0.1 H$_2$O maser gives $\tau_{\rm dyn}=750 \pm 100$ yr for the north-south bipolar extension, which is not much different from those of massive star forming regions, for instance, $\tau_{\rm dyn}\sim 500$ for W49N (Gwinn et al.1992).
On the other hand, H$_2$O maser motion in these massive star forming regions exceed 100 km s$^{-1}$, while that in OH 43.8$-$0.1 is $\sim$ 8 km s$^{-1}$.
Thus, this scale difference reflects the difference of outflow energy between OH 43.8$-$0.1 and other massive star forming regions, and this is presumably linked to lower power, implying lower mass of a forming star in OH 43.8$-$0.1 than the protostars in W49N, Sgr B2 and Orion-KL.

\subsection{Constraints on Galactic Rotation Curve}

Although this study presents a distance measurement of only one source, the result can be used to constrain the rotation properties of the Galaxy.
Assuming a circular rotation at any radius of the Galaxy, observed radial velocity can be written as,
\begin{equation}
V_r = \Bigl[ \Theta (R)\left(\frac{R_0}{R}\right) - \Theta_0 \Bigr] \sin l.
\label{eq:V_r}
\end{equation}
Here $R$ and $R_0$ are the galacto-centric distances of the source (OH 43.8$-$0.1 for the present study) and the Sun, and $\Theta (R)$ and $\Theta_0$ are circular rotation speed of the Galaxy at the radius of the source and the Sun, and $l$ is the galactic longitude.
Note that $V_r$ and $l$ are observables, and $R$ can also be regarded as an observable because $R$ is written as
\begin{equation}
R = \sqrt{D^2 + R_0^2 - 2D R_0 \cos l},
\label{eq:R}
\end{equation}
and hence one can immediately know $R$ once $D$ and $R_0$ are given.
Equation (\ref{eq:V_r}) contains three parameters to be determined, namely, $R_0$, $\Theta_0$, $\Theta (R)$.
Among them, the galactic constants, $R_0$ and $\Theta_0$, are not necessarily unknown as there have been numerous studies to evaluate them.
Therefore, if $R_0$ and $\Theta_0$ are given, then $\Theta (R)$ can be determined, giving a constraint on the galactic rotation speed.

The IAU standard values are $R_0=8.5$ kpc and $\Theta_0=220$ km s$^{-1}$ (Kerr \& Lynden-Bell 1986), but several recent studies claim different values.
Determinations of $R_0$ later than 1986 tend to give smaller values, and Reid (1993) summarized the determinations of $R_0$, giving $R_0=8.0 \pm 0.5$ kpc.
As for $\Theta_0$ the situation is much complicated: while number of recent studies with various methods gave values around 200 $\sim$ 220 km $^{-1}$, being in good agreement with previous ones (e.g., Mignard 2000; Bedin et al. 2003; Kalirai et al. 2004; Reid and Brunthaler 2004), there were also some studies which report fairly smaller or larger values varying from 180 to 270 km s$^{-1}$ (e.g., Olling \& Merrifield 1998; Miyamoto \& Zhu 1998).
In the following analyses, we conservatively take large error for both $R_0$ and $\Theta_0$, taking $R_0=8.0 \pm 1.0$ kpc and $\Theta_0=220 \pm 50$ km s$^{-1}$.

Here we rewrite equation (\ref{eq:V_r}) by introducing a new parameter $\theta_r \equiv \Theta(R)/\Theta_0$ as,
\begin{equation}
\theta_r = \left( \frac{V_r}{\Theta_0 \sin l}+1 \right) \frac{R}{R_0}.
\label{eq:theta0}
\end{equation}
Based on the galactic constants $R_0$ and $\Theta_0$, one can determine $\theta_r$, which is the ratio of the galactic rotation speeds at the source and at the Sun.
If the rotation curve is completely flat as found in many bright spiral galaxies (e.g., Sofue \& Rubin 2000), $\theta_r$ should be unity.
In order to obtain $\theta_r$ from equation (\ref{eq:theta0}), one also have to know the value of $V_r$, the radial velocity.
Assuming that the central cluster of H$_2$O maser is indeed associated with the UC HII regions and thus its radial velocity reflect the systematic velocity of OH 43.8$-$0.1, the averaged velocity give $V_{\rm LSR}=40$ km s$^{-1}$.
Observations of molecular gases with CS and CO lines give $V_{\rm LSR}=43 \sim 44$ km s$^{-1}$ (Plume et al. 1992; Shepherd \& Churchwell 1996), and methanol maser lines and OH maser lines lie in the LSR velocity range of 39 to 44 km $s^{-1}$ (Menten 1991; Szymczak et al. 2000; Argon et al. 2000).
Summarizing these results, here we adopt $V_r=42 \pm 3$ km $s^{-1}$ as the systemic velocity of OH 43.8$-$0.1.
Using these values ($R_0=8.0 \pm 1.0$ kpc, $\Theta_0=220 \pm 50$ km s$^{-1}$, $D=2.8 \pm 0.5$ kpc, and $V_r=42 \pm 3$ km s$^{-1}$), equation (\ref{eq:theta0}) yields that
\begin{equation}
\theta_r = 1.00\; ^{+0.10}_{-0.06},
\end{equation}
where the upper and lower limits are calculated as the worst values with possible combinations of above parameter ranges.
The value of $\theta_r$ indicates that there is no systematic difference between rotation speeds at OH 43.8$-$0.1 and at the Sun, and thus that the galactic rotation curve is fairly flat, being consistent with previous studies (e.g., Honma \& Sofue 1997).
For $R_0=8.0$ kpc, equation (\ref{eq:R}) gives $R=6.3$ kpc as a galacto-centric distance of OH 43.8$-$0.1, and the interpolation of rotation speed between the Sun and OH 43.8$-$0.1 suggests that flat rotation curve spans nearly 2 kpc.

\section{Summary and Concluding Remark}

We carried out 5-epoch observations of OH 43.8$-$0.1 H$_2$O maser with VERA, and successfully measured the proper motions of OH 43.8$-$0.1 H$_2$O masers for the first time.
The large-scale maser map revealed north-south extension of maser spots, and proper motion measurements showed that the distribution as well as the kinematics of OH 43.8$-$0.1 H$_2$O maser can be represented by the bipolar flow in north-south direction plus the central cluster which has rather complex velocity structure.
Based on statistical parallax, we also determined the distance to OH 43.8$-$0.1 as 2.8 $\pm$ 0.5 kpc, ruling out a far kinematic distance.
The distance, combined with previous studies on the galactic constants such as $R_0$ and $\Theta_0$, gave a constraint on the flatness of galactic rotation curve, showing no systematic difference in the galactic rotation velocity at the Sun and OH 43.8$-$0.1.

In the near future, VERA will observe hundreds of H$_2$O maser sources to perform astrometry of galactic star forming regions.
While the main target of VERA is to investigate the structure and dynamics of the Milky Way Galaxy via astrometry of maser sources, VERA will also provide useful information on internal structures and kinematics of star forming regions, and the present study is one of the first examples of such valuable by-products of VERA.

\vspace{1pc}\par
One of the authors (MH) acknowledges financial support from Inamori Foundation and from grant-in-aid (No.16740120) from the Ministry of Education, Culture, Sports, Science and Technology (MEXT).
Part of the data reduction was performed at the Astronomical Data Analysis Center (ADAC) of the NAOJ, which is an inter-university research institute of astronomy operated by MEXT.


\clearpage

\begin{table}
\footnotesize
Table~1. \hspace{4pt} Results of proper motion measurements.\\
\begin{tabular}{rrrrrrcrr}
\hline
 ID & $V_{\rm LSR}$ &  $X\;\;\;\;\;$ & $Y\;\;\;\;\;$ &  $\mu_X\;\;$ & $\mu_Y\;\;$ & epoch & $\delta_X\;\;$ & $\delta_Y\;\;$ \\
\hline
  1 & 50.92 &$-$145.105 & 136.124 &  0.129 &  0.152 & 00111 & 0.012 & 0.027 \\
  2 & 50.47 &$-$145.085 & 136.030 &  0.088 &  0.416 & 11111 & 0.028 & 0.035 \\
  3 & 50.04 &$-$145.132 & 136.030 &  0.202 &  0.461 & 11111 & 0.031 & 0.026 \\
  4 & 49.62 &$-$145.117 & 135.997 &  0.180 &  0.529 & 11111 & 0.030 & 0.028 \\
  5 & 49.20 &$-$145.102 & 135.982 &  0.180 &  0.552 & 11111 & 0.017 & 0.030 \\
  6 & 48.78 &$-$145.092 & 135.961 &  0.146 &  0.689 & 11111 & 0.016 & 0.024 \\
  7 & 48.36 &$-$145.089 & 135.957 &  0.146 &  0.740 & 11111 & 0.021 & 0.016 \\
  8 & 47.94 &$-$145.093 & 135.998 &  0.144 &  0.699 & 11111 & 0.020 & 0.036 \\
  9 & 47.52 &$-$145.084 & 136.015 &  0.140 &  0.647 & 11111 & 0.015 & 0.022 \\
 10 & 47.10 &$-$145.076 & 136.030 &  0.166 &  0.570 & 11111 & 0.011 & 0.023 \\
 11 & 46.67 &$-$145.090 & 136.040 &  0.197 &  0.518 & 11111 & 0.026 & 0.039 \\
 12 & 46.29 &$-$155.368 & 134.127 &  0.088 &  0.325 & 10101 & 0.035 & 0.029 \\
 13 & 45.02 & $-$82.968 &$-$276.683 & $-$0.011 &  0.204 & 00111 & 0.000 & 0.058 \\
 14 & 44.99 &$-$267.836 &$-$507.013 &  0.113 & $-$0.212 & 11101 & 0.033 & 0.069 \\
 15 & 44.56 & $-$82.884 &$-$276.457 & $-$0.159 & $-$0.229 & 01111 & 0.031 & 0.066 \\
 16 & 44.15 & $-$82.811 &$-$276.482 & $-$0.281 & $-$0.279 & 11111 & 0.031 & 0.047 \\
 17 & 43.72 & $-$82.815 &$-$276.430 & $-$0.286 & $-$0.509 & 11111 & 0.021 & 0.053 \\
 18 & 43.29 & $-$82.809 &$-$276.490 & $-$0.245 &  0.037 & 11100 & 0.042 & 0.019 \\
 19 & 42.87 &$-$155.561 & 134.052 &  0.562 &  0.391 & 11100 & 0.004 & 0.064 \\
 20 & 42.49 &$-$218.197 & 104.703 & $-$0.117 & $-$0.039 & 00111 & 0.017 & 0.006 \\
 21 & 42.07 &$-$218.168 & 104.679 & $-$0.181 &  0.022 & 00111 & 0.005 & 0.002 \\
 22 & 41.65 &$-$218.141 & 104.652 & $-$0.264 &  0.103 & 00111 & 0.007 & 0.012 \\
 23 & 41.62 &$-$271.528 & $-$48.440 & $-$0.097 & $-$0.117 & 11111 & 0.023 & 0.032 \\
 24 & 41.20 & 128.780 &$-$809.712 &  0.454 & $-$0.599 & 11111 & 0.042 & 0.078 \\
 25 & 40.76 &$-$215.215 & 106.730 &  0.131 &  0.014 & 11100 & 0.020 & 0.016 \\
 26 & 40.38 & $-$75.557 &  73.994 &  0.221 &  0.100 & 00111 & 0.049 & 0.011 \\
 27 & 40.23 & 398.851 &$-$106.395 &  0.553 & $-$0.164 & 01111 & 0.008 & 0.082 \\
 28 & 39.96 & $-$51.367 &  44.381 &  0.102 &  0.176 & 00111 & 0.017 & 0.008 \\
 29 & 39.96 & $-$18.276 &$-$310.044 & $-$0.297 & $-$0.056 & 00111 & 0.028 & 0.012 \\
 30 & 39.51 & $-$51.386 &  44.441 &  0.163 &  0.116 & 11111 & 0.010 & 0.022 \\
 31 & 39.51 & $-$18.438 &$-$310.014 & $-$0.127 & $-$0.103 & 11111 & 0.006 & 0.015 \\
 32 & 39.09 & $-$18.481 &$-$310.026 & $-$0.018 & $-$0.104 & 11111 & 0.013 & 0.010 \\
 33 & 38.66 & $-$18.555 &$-$309.907 &  0.130 & $-$0.297 & 01111 & 0.010 & 0.014 \\
 34 & 38.66 &  $-$0.008 &   0.022 &  0.004 & $-$0.078 & 11110 & 0.003 & 0.006 \\
 35 & 38.25 &   0.000 &   0.000 &  0.000 &  0.000 & 11111 & 0.000 & 0.000 \\
 36 & 37.85 & $-$65.069 &$-$250.810 & $-$0.046 & $-$0.317 & 00111 & 0.010 & 0.005 \\
 37 & 37.81 &  $-$0.023 &   0.029 &  0.025 & $-$0.017 & 11100 & 0.019 & 0.009 \\
 38 & 37.40 & $-$65.013 &$-$250.881 & $-$0.128 & $-$0.094 & 11101 & 0.001 & 0.048 \\
 39 & 36.13 & $-$52.904 &$-$249.517 &  0.822 & $-$0.185 & 11110 & 0.047 & 0.024 \\
 40 & 35.72 & $-$52.870 &$-$249.514 &  0.809 & $-$0.218 & 11111 & 0.054 & 0.040 \\
 41 & 35.30 & $-$52.734 &$-$249.572 &  0.576 & $-$0.064 & 11111 & 0.048 & 0.032 \\
 42 & 34.88 & $-$52.602 &$-$249.609 &  0.392 & $-$0.012 & 11111 & 0.034 & 0.024 \\
 43 & 34.44 & $-$52.616 &$-$249.555 &  0.443 & $-$0.058 & 00111 & 0.060 & 0.023 \\
\hline\\
\end{tabular}
\\
Column 1: spot ID number.\\
Column 2: LSR velocity in km s$^{-1}$ (averaged over detected epochs).\\
Column 3 \& 4: best fit positions at the first epoch (in mas). \\
Column 5 \& 6: best-fit proper motions (in mass).\\
Column 7: detected epoch. {`}1{'} for detection, and {`}0{'} for non-detection.\\
Column 8 \& 9: RMS residuals from best fit proper motions (in mass).\\

\end{table}

\end{document}